\documentstyle[sprocl]{article}

\input epsf

\def\be{\begin{equation}}
\def\ee{\end{equation}}
\def\bea{\begin{eqnarray}}
\def\eea{\end{eqnarray}}

\begin{document}

\title{WHAT WE KNOW ABOUT THE THEORETICAL FOUNDATION OF DUALITY IN
ELECTRON SCATTERING}

\author{ C. E. CARLSON }

\address{Nuclear and Particle Theory Group, Physics Department\\
College of William and Mary, Williamsburg, VA 23187-8795, USA}


\maketitle\abstracts{
We consider some of the things that we understand about the theoretical
underpinnings of duality, including items such as why the
resonance peak/background ratio is constant in general, why it falls for
the $\Delta(1232)$, what we might expect for scaling and duality with
longitudinal or spin-dependent structure functions, and what kind of
scaling or duality we might expect for semiexclusive processes.
}


\section{Introduction}

One has to confess that we don't understand the theoretical foundation of
duality in electron scattering.  That should not place a total damper on
this discussion.  We do understand some things, and we will try to
explain what we do understand.  

In this talk, after some preliminary remarks defining what we mean by
duality in electron (or more generally, lepton) scattering, and some
further preliminary remarks about why it could be useful to understand
duality well, we will examine a number of more specific topics where we
do have some understanding or can make predictions, including

\begin{itemize}

\item  Why, in general, do the resonance peaks stick up above the 
smooth ``background'' curve by the same ratio regardless of the
momentum transfer involved?

\item Why would a specific resonance like the $\Delta(1232)$ be an
exception to the above, or, why does the $\Delta(1232)$ disappear?

\item What kind of scaling or duality do we expect for the longitudinal
structure function?

\item What kind of scaling or duality do we expect for data with
polarized initial states?

\item What kind of scaling or duality can we look for with
semi-exclusive data?

\end{itemize}

We will continue this introduction with the promised remarks, and then in
section~\ref{sectiontwo} answer as well as we can the questions posed.

\subsection {Statement of duality}

One can separate two aspects of duality.  One is ``local duality,''
and the other the the constancy of the resonance peak to background ratio,
which can be viewed as discussing how the local duality is realized. 
``Duality,'' if unqualified is often taken to mean local duality.

Local duality, if there if no evolution of the structure function,  is
constancy of an average of the structure function over a limited
$x$ region. Take the brackets
$\langle \ldots \rangle$ to mean an average over a region of $x$ that can
include some chosen resonance at low $Q^2$.  Then duality implies
equality between
$\langle F_2(x,Q^2) \rangle$ evaluated at a low
$Q^2$, where $F_2$ is in the resonance region, to the same quantity and
the same
$x$ region but at a high $Q^2$ in the scaling region.

If there is evolution, one would think one should
compare $\langle F_2(x,Q^2) \rangle$ for real data at low $Q^2$, in
the resonance region, to the same quantity and the same $x$ region but for
the smooth scaling curve evolved to the same $Q^2$.
There is, however, an interesting and unsettled question that we won't
discuss, and that is whether there is reduced
$F_2$ evolution in the resonance region~\cite{nonevolution}.

Broadly, the appearance of duality in the data tells us that the single
quark reaction rate determines accurately the reaction rate for the entire
process, including final state interactions---on the average.

\subsection{Possible uses of duality }

There are useful experimental studies we could undertake if the role of
the final state interactions in forming the resonance becomes moot when
averaged over, say, the resonance width.  If reliably understood, 
duality could be useful.

\begin{itemize}

\item One could study the structure functions in the $x
\rightarrow 1$ region.  For a fixed available energy, $x \rightarrow 1$
means getting into the resonance region and if one were sure of the
connection of the resonance region average to the scaling curve, one
could determine the scaling result for $F_2$ significantly closer to the
kinematic upper endpoint.

\item Similar remarks for apply to the semiexclusive reaction, 
$\gamma^*(q)  + p \rightarrow \pi(k) + X$, with the pion emerging with
3-momentum parallel to that of the virtual photon, as proposed by C.
Armstrong {\it et al}.

\end{itemize}

We will now proceed to a few more specific points about what is
known and what could further be studied in exclusive-inclusive
connections in electron scattering.  We should remark that there
has been a ``proof'', or at least a ``demystification,'' of duality
in an interesting paper by DeR\'ujula, Georgi, and Politzer in
1977~\cite{dgp77}.  There seems, however, room for more discussion.


\section{Discussion}                           \label{sectiontwo}

\subsection{Why constant signal/background or resonance/continuum (in
general)?}

\begin{figure} 

\hskip .1in\epsfxsize 4.3in \epsfbox{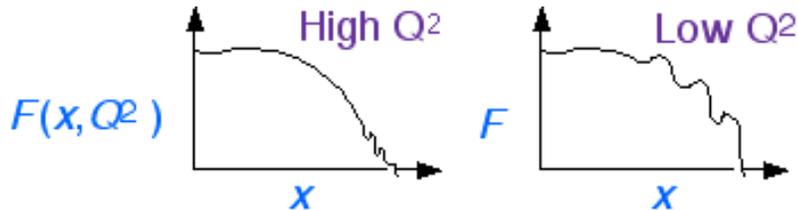}

\caption{Scaling curves, with locations of resonance bumps at low and
high $Q^2$.}

\end{figure}

There always is a resonance region.  As
$Q^2$ increases it slides closer to the kinematic endpoint $x = 1$. 
We may ask, using the language that the ``signal'' is the resonance peak
``continuum'' is gotten from the scaling curve, whether or not we expect
the signal to continuum  ratio to be constant---as it often appears to
be in Nature?

Note that this question is logically independent of local
duality.  One way to realize local duality is allows a constant signal
to continuum ratio.  But local duality could also be realized by having
the resonance disappear into the background beneath it, with the total
averaging out in a way that preserves local duality.  Conversely, the
resonance peak could be very large, with the average unequal to what is
obtained from the scaling curve, and yet the signal/continuum ratio be
constant.

To return to the question, it does appear we can prove that
within perturbative QCD we do expect a constant signal to
continuum ration~\cite{cm90}.  The proof has one requirement each on the
scaling curve and the resonance production form factors, namely

\begin{itemize}
\item  a $(1-x)^3$ behavior for scaling curve as $x\rightarrow 1$ (which
can itself be proved in perturbative QCD), and
\item pQCD scaling (in $Q^2$) of the leading (helicity conserving)
resonance form factor
\end{itemize}

The proof was written out in the live version of these notes, and may be
examined in~\cite{cm90}.  Basically, it proceeds by writing the cross
section for production of a resonance with finite width, switching
variables from $Q^2$ (for example, in the nucleon to resonance transition
form factors) to $x$, and then comparing the result to the deep inelastic
cross section to recognize the connection between $F_2$ and the form
factors. The result works for most known resonances.

\subsection{Why the Delta(1232) disappears}

The $\Delta(1232)$, unlike resonances in the 1535 or 1688 MeV regions,
becomes progressively harder to find as $Q^2$ increases.  This violates
the theorem whose proof was just outlined, and one would like to know why.

First, however, let us point out that local duality is still maintained.
Duality means that the average over the resonance region matches
the the average using the scaling curve.  It does--even for
$\Delta(1232)$.  What happens is that as the resonance peak falls, the
background rises, and average/continuum $\approx$ const.~\cite{cm93}

One concludes that the background knows about the $\Delta$, and 
co-concludes that one should not use just simple $\pi$-nucleon Born
terms to model the background.

Why $\Delta$ disappears is simply because the asymptotic size of
the leading helicity form factor is anomalously small.  This is not
just a result of observation, but also a result of a pQCD calculation,
similar to the better known pQCD calculation of the high
$Q^2$ nucleon elastic dirac form factor
$F_1$.

\begin{figure}

\centerline {\epsfxsize 2.3 in \epsfbox{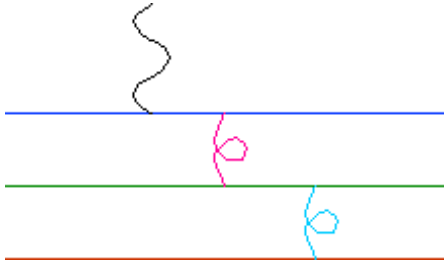}}

\caption{Example of lowest order diagram that underlies pQCD
form factor calculation.}

\end{figure}

Hence what we mainly see in $N \rightarrow \Delta$ are
asymptotically subleading amplitudes. It is a lousy circumstance for
pQCD that the first resonance is an exceptional case, yet it is a
circumstance substantiated by calculation.

\subsection{The longitudinal structure function}

We will just assert that we still expect duality to work.  In particular,
we expect the signal to continuum ratio is the same at all $Q^2$, just
as it is for $F_2$, which is dominantly the transverse structure
function.  This assertion follows a prediction of pQCD, this time
given~\cite{cm90}

\begin{itemize}
\item a $(1-x)^4$ behavior for the longitudinal structure function scaling
curve as
$x\rightarrow 1$ (again itself a result of pQCD), and
\item pQCD scaling (in $Q^2$) of resonance form factor for longitudinal
photons (one quark helicity flip)
\end{itemize}

\noindent But there may be some differences.  For example, 
\begin{itemize}                                   
\item  The signal to continuum ratio may be constant even for the
$\Delta(1232)$ That the leading helicity amplitude is anomalously small
does not mean that the next-to-leading helicity
amplitude (which is not currently calculable in pQCD) is also
small.  If it is normal size, the $\Delta(1232)$ will not be a
disappearing resonance in the longitudinal channel.  
\item Maybe the Roper, the $N(1440)$, will appear.  It has not been
observed in electroproduction when measuring the transverse channel.
There is an interesting possibility that {\it if} the Roper is hybrid
baryon (meaning its lowest significant Fock component is a qqqg state),
its leading electroproduction amplitude is asymptotically
$1/Q^2$ smaller than qqq, but its longitudinal amplitude has normal
falloff~\cite{cm91}.
\end{itemize}

\subsection{Expectation with polarized initial states}

We expect duality to work for spin dependent structure function, $g_1$,
also.  Time was, and space is, short, and so we shall just
refer to reference~\cite{cm98} and the work of X. Ji {\it et
al.}~\cite{ji}.


\subsection{What to look for with semi-exclusive data}

\begin{figure}

\centerline{ \epsfysize 1.8in  \epsfbox{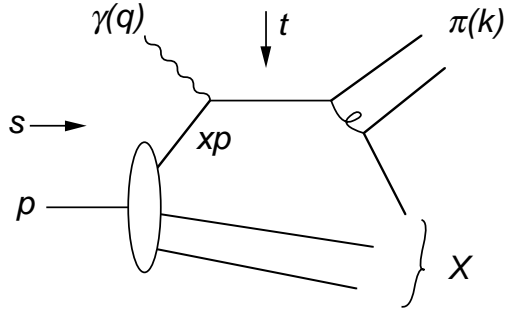}  }

\caption{Direct or short distance pion production.}

\label{direct}

\end{figure}

In 
$$\gamma(q)  + p \rightarrow \pi(k) + X, $$
if the pion is produced in a 
direct or short range process, illustrated in Fig.~\ref{direct}, then we
can show that 
there will be a function~\cite{acw00}, $F(x,s,t,q^2)$, for which
there will be a scaling region where it is dependent mainly on $x$,
\begin{eqnarray}
x \equiv {-t \over s + u - 2 m_N^2 - q^2 - m_\pi^2}  ,
\nonumber
\end{eqnarray}
where $s$, $t$, and $u$ are Mandelstam variables, and---at least for
the direct process---$x$ is the momentum fraction of the struck
quark~\cite{acw,bdhp}, just as in deep inelastic scattering.

For scaling, one needs $s$, $t$, $u$, and $m_X$ large.
We can get $m_X$ into the resonance region with fixed $q^2$  and
diminishing $t$.  Will we see an inclusive-exclusive connection as in the
DIS case?

\begin{figure}

\vglue .1in \hglue -.1in     \epsfxsize 2.2in \epsfbox{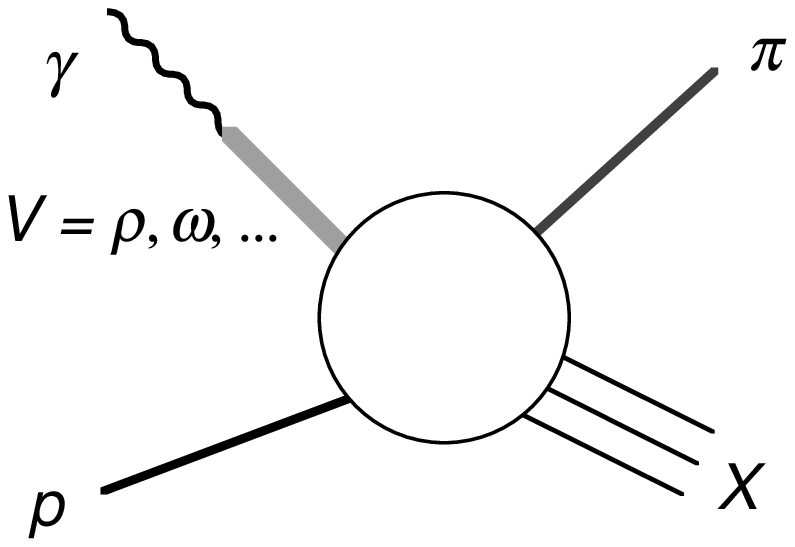}  

\vglue -1.4in \hglue 2.3in \epsfxsize 2.2in \epsfbox{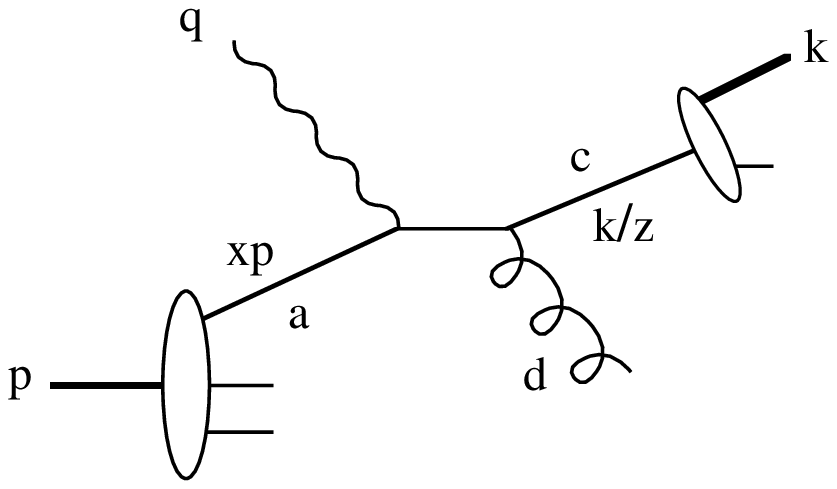}

\caption{A vector meson dominated and a fragmentation process.}

\label{background}

\end{figure}

First we have to see scaling, which in addition to requiring large values
of the kinematic variables also requires that the competing processes,
such as the soft or vector meson dominated (VMD) process and fragmentation
(illustrated in Fig.~\ref{background}), be small.

\begin{figure}

\epsfxsize 4.7 in \epsfbox{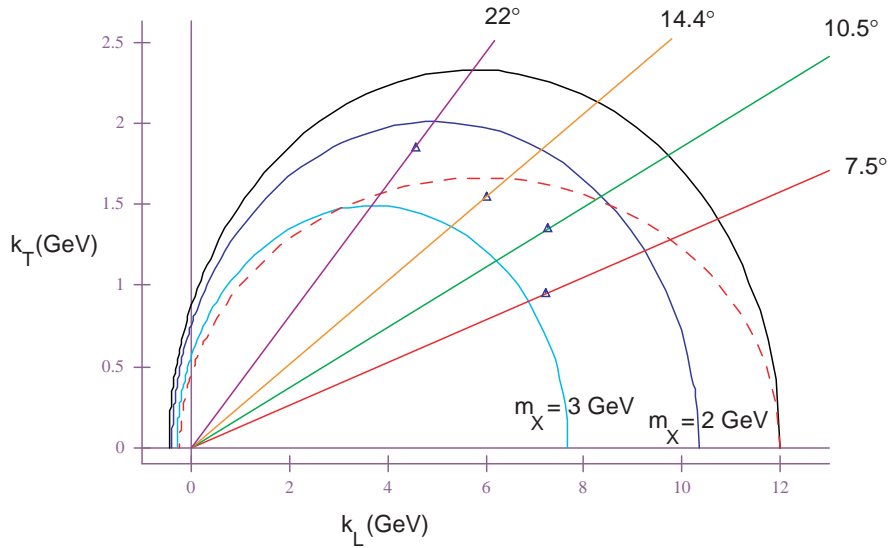}  

\caption{Kinematic regions for $\gamma(q)  + p \rightarrow \pi^+(k) + X$.
Direct or short distance processes dominate above and to the right of
the small triangles.  The solid ellipses show $m_X = m_N$, 2 GeV, 3 GeV.
The dashed ellipse is for fixed $x = 0.5$.  }

\label{dualregions}

\end{figure}

VMD is  a serious background for photoproduction with 12 GeV photons. 
We can decrease the size of VMD process by using spacelike off-shell
photons, rather than real photons, since
\begin{eqnarray}
{1\over m_\rho^2} \rightarrow {1\over Q^2 + m_\rho^2} \nonumber
\end{eqnarray}
We have from earlier work the means to calculate direct process~\cite{acw}
and estimate VMD process~\cite{acw99}.  A preliminary result for
$\gamma(q)  + p \rightarrow \pi^+(k) + X $ is shown
in Fig.~\ref{dualregions} for electroproduction where we have chosen
the incoming photons to have and energy of 12 GeV and to be off-shell by
1 GeV$^2$ spacelike.  The straight lines show several different lab
angles for the outgoing pion, and the small triangles are crucial marks. 
Above and to the right of the triangles, direct or short distance pion
production dominates over VDM or fragmentation processes, so that in this
region we can connect an experimentally measurable $x$ to the momentum
fraction of the struck quark, define a scaling function, and then follow
what happens as we enter the resonance region.  The solid curved lines,
which one can show are ellipses, show the boundaries for $m_X$ being
$m_N$, 2 Gev, and 3 GeV.  The region between the outer two
curves is the resonance region, and inside the $m_X = 2$ GeV curve is the
scaling region.  There is also one curve, a dashed ellipse, showing the
path of constant $x = 0.5$ in this diagram.


Thus the kinematics exists for allowing  a study of some function 
$F(x,s,t,q^2)$ which should scale in the region $m_X > 2$ GeV and high
$Q^2$, and one can see if it too exhibits the properties of local duality
and constant signal to continuum ratio as one enters the resonance region.

\section*{Acknowledgments}
I cannot think about duality without thinking of Nimai Mukhopadhyay and
the happy times we spent discussing this and other subjects.  He has my
deepest gratitude.

Thanks to the organizers for an excellent conference, also to the National
Science  Foundation for support under Grant No.\ PHY-9900657.

\end{document}